\documentclass[10pt, sigconf]{acmart}
\makeatletter
\def\@ACM@checkaffil{
    \if@ACM@instpresent\else
    \ClassWarningNoLine{\@classname}{No institution present for an affiliation}%
    \fi
    \if@ACM@citypresent\else
    \ClassWarningNoLine{\@classname}{No city present for an affiliation}%
    \fi
    \if@ACM@countrypresent\else
        \ClassWarningNoLine{\@classname}{No country present for an affiliation}%
    \fi
}
\makeatother
\usepackage{hyperref}

\usepackage[english]{babel}
\usepackage{blindtext}

\usepackage{algorithm}
\usepackage[noend]{algpseudocode}

\usepackage{graphicx}
\usepackage{newtxmath}
\usepackage{lipsum}
\usepackage{subfigure}
\usepackage{textcomp}
\usepackage{xcolor}
\usepackage{bm}
\usepackage{url}
\usepackage{tabularx,booktabs}
\usepackage{tabularray}

\renewcommand\footnotetextcopyrightpermission[1]{} 
\setcopyright{none}

\newif\ifcolormarker
\colormarkertrue

\newcommand\blfootnote[1]{%
  \begingroup
  \renewcommand\thefootnote{}\footnote{#1}%
  \addtocounter{footnote}{-1}%
  \endgroup
}

\acmConference[GNNet '23] {Proceedings of the 2nd Graph Neural Networking Workshop 2023}{December 8, 2023}{Paris, France.}

\begin{document}

\title[Atom: Neural Traffic Compression with STGNNs]{Atom: Neural Traffic Compression with Spatio-Temporal Graph Neural Networks}

\author{Paul Almasan\footnotemark[1]$^{\diamond}$, Krzysztof Rusek\footnotemark[2], Shihan Xiao\footnotemark[3], Xiang Shi\footnotemark[3], Xiangle Cheng\footnotemark[3],\\Albert Cabellos-Aparicio\footnotemark[4], Pere Barlet-Ros\footnotemark[4]}
\thanks{$\diamond$ Work undertaken during a PhD program at Universitat Politècnica de Catalunya.}

\affiliation{%
  \institution{
  \footnotemark[1]Telefónica Research
  \footnotemark[2]AGH University of Science and Technology, Institute of Telecommunications\\
  \footnotemark[3]Network Technology Lab., Huawei Technologies Co., Ltd.\\\vspace{-0.1cm}
  \footnotemark[4]Barcelona Neural Networking Center, Universitat Polit\`ecnica de Catalunya\\
  }
}

\renewcommand{\shortauthors}{Paul Almasan et al.}

\begin{abstract} 
Storing network traffic data is key to efficient network management; however, it is becoming more challenging and costly due to the ever-increasing data transmission rates, traffic volumes, and connected devices. In this paper, we explore the use of neural architectures for network traffic compression. Specifically, we consider a network scenario with multiple measurement points in a network topology. Such measurements can be interpreted as multiple time series that exhibit spatial and temporal correlations induced by network topology, routing, or user behavior. We present \textit{Atom}, a neural traffic compression method that leverages spatial and temporal correlations present in network traffic. \textit{Atom} implements a customized spatio-temporal graph neural network design that effectively exploits both types of correlations simultaneously. The experimental results show that \textit{Atom} can outperform GZIP's compression ratios by 50\%-65\% on three real-world networks.
\end{abstract}

\maketitle

\section{Introduction}
\blfootnote{If you cite this paper, please use the CoNEXT '23 reference:
Paul Almasan, Krzysztof Rusek, Shihan Xiao, Xiang Shi, Xiangle Cheng, Albert Cabellos-Aparicio, Pere Barlet-Ros. 2023. Atom: Neural Traffic Compression with Spatio-Temporal Graph Neural Networks. In Proceedings of the 2nd International Workshop on Graph Neural Networking (GNNet '23). Association for Computing Machinery, New York, NY, USA. \url{https://doi.org/10.1145/3630049.3630170}}

In recent years, modern networks have experienced a significant surge in both network traffic and connected devices~\cite{tune2013internet, alam2018reliable}. This growth has been fueled by the widespread deployment of advanced applications such as vehicular networks, IoT, virtual reality, video streaming, and Industry 4.0. Simultaneously, ongoing advancements in network technology such as improved link speeds have intensified this upward trajectory. To manage these networks effectively, network operators must store network traffic information, including packet traces, link-level traffic measurements, and flow-level measurements. This information plays a crucial role in various network management tasks, including network planning, traffic engineering, traffic classification, anomaly detection, and network forensics. In addition, the emerging concept of network digital twins requires the storage and analysis of vast volumes of network traffic data~\cite{9374645, 9795043}.

The efficient storage of network traffic has become increasingly challenging due to the exponential growth in data volume. Traffic traces generated by Internet Service Providers (ISPs), backbone networks, mobile networks, or data centers can easily consume hundreds of terabytes or petabytes per day~\cite{10.1145/2829988.2787472,alam2018reliable}.
This problem is exacerbated because current network topologies can have hundreds of links~\cite{knight2011internet}.

Typically, network operators use generic lossless compression methods such as GZIP~\cite{gzip_reference} to compress network traffic data. However, traditional compression methods may not deliver optimal compression performance when applied to network-traffic data. 
Previous studies have consistently demonstrated that network traffic has an inherent structure, refuting the notion that is purely random ~\cite{tune2013internet, 10.1145/1879141.1879175, lan2006measurement, gao2020predicting}. Consequently, network traffic exhibits discernible spatial and temporal patterns that offer opportunities for improving compression ratios. 

In this work, we aimed to investigate whether recent advancements in Neural Network (NN) architectures can effectively harness the spatial and temporal correlations present in network traffic to achieve better compression ratios compared to conventional tools such as GZIP. To this end, we considered a scenario that involves the compression of network traffic data obtained using standard tools such as SNMP or NetFlow from multiple vantage points. We expected the data collected from multiple measurement points within a network topology to exhibit spatio-temporal correlations. 

In this paper, we first introduce \textit{Atom}, a neural traffic compression method that exploits spatial and temporal correlations naturally present in network traffic. Our compression method combines a neural network with Arithmetic Coding (AC) \cite{10.1145/214762.214771} to effectively compress link-level traffic measurements. Second, we describe a custom Spatio-Temporal Graph Neural Network (ST-GNN) used to exploit spatial and temporal correlations. Finally, we evaluate \textit{Atom}'s compression performance by conducting comprehensive experiments on synthetic and real-world datasets.

The experimental results show that \textit{Atom} can improve GZIP's compression ratios by 35\%-95\% on synthetic data with different levels of spatial and temporal correlation. When compressing real-world data, the experimental results indicate that \textit{Atom} can reduce the data size by 50\%-65\% compared to GZIP, and by a factor between 2.6x and 4.2x with respect to the original file. The source code of \textit{Atom} together with the datasets are publicly available\footnote{\url{https://github.com/BNN-UPC/Atom_Neural_Traffic_Compression}}.

\section{Methodology} 

\begin{figure}[!t]
  \centering
  \includegraphics[width=0.8\linewidth]{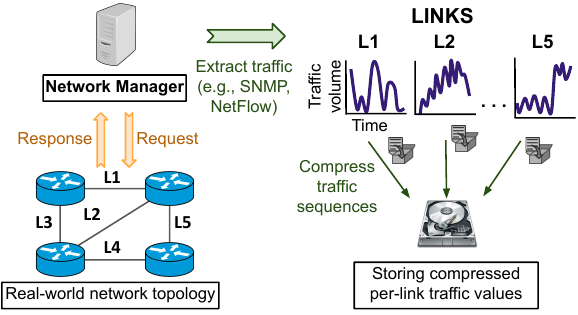}
  \caption{Overview of the network traffic compression scenario. 
  }
  \label{fig:compress_scenario}
\vspace{-0.4cm}
\end{figure}

We consider a network scenario with multiple vantage points within a network topology (e.g., one per link), where traffic measurements are preformed periodically over time using standard measurement tools, such as SNMP or NetFlow. Traffic data is stored in time bins (e.g., bins of 5 minutes), resulting in a sequence of traffic data that we want to compress. Figure~\ref{fig:compress_scenario} shows an overview of the compression scenario.
The metric we use to compare the compression performance of different methods is the Compression Ratio (CR). This ratio is computed by dividing the uncompressed file size by the compressed file size, returning a value that indicates how much smaller is the file size after compression.

We consider two different compression scenarios. In the first scenario, we want to compress the traffic values from a single measurement point independently (e.g., access link). This is a common practice in small- or medium-sized networks, where internal traffic is smaller and is not considered to be of interest in many cases (e.g., enterprise, or campus networks). The second scenario represents a more general use case, where we want to simultaneously compress the traffic from multiple measurement points of a network topology (i.e., network-wide compression). This situation is more common in large networks, such as ISPs or mobile networks, which can have a global view of the network topology. 

\subsection{Arithmetic Coding}
\label{subsec:arithm_coding}

\textit{Atom} leverages arithmetic coding \cite{10.1145/214762.214771} to implement the \textit{encoder/decoder} (see Section~\ref{subsec:encoddecod}). AC is a lossless method that compresses a stream of symbols (e.g., text characters) into a single number between [0, 1). This is done by assigning fewer bits to frequent symbols and more bits to less frequent symbols. In contrast to other popular compression methods such as Huffman coding~\cite{4051119}, AC achieves better compression ratios and it can work in an online fashion. In addition, AC works with probability distributions, making it a good fit with NN technologies.

The compression performance of AC is defined by the quality of the probability distribution. In other words, if the predictor is accurate, AC will assign fewer bits to encode the symbol, resulting in close-to-optimal compression performance. On the other hand, if the predictor is not accurate, the probabilities will not correspond to the real symbol, which results in poor compression or it can even increase the final file size.

\subsection{Notation and problem statement}
\label{subsec:notation}
Formally, traffic measurements are represented as a matrix $\bm X\in \mathbb{N}^{w\times l}$, where $w$ represents the Sliding Window (SW) for $l$ links.
Each traffic measurement is a random vector $\bm x_t\in\mathbb{N}^{l}$.
For the arithmetic encoder we need a one step forecast distribution $p(\bm x_{t}| \bm x_{<t})$ to capture temporal dependence.
As the arithmetic encoder operates on streams of symbols, we further partition the distribution with a chain rule to capture spatial dependence:
\begin{equation}
    p(\bm x_{t}|\bm x_{<t}) = \prod_l p(x_l|\bm x_{t,<l},\bm x_{<t}).
 \label{fig:spatial_equation}
\end{equation}
Here we assume the stationary model $p(x_l|\bm x_{t,<l},\bm x_{<t})$ and mask-out the unknown traffic values.
Note that there is no natural order for the auto-regressive model, however, the only requirement is that the order must be the same during compression and decompression.

\section{Design}
\label{sec:design}

\begin{figure}[!t]
  \centering
  \includegraphics[width=0.85\linewidth]{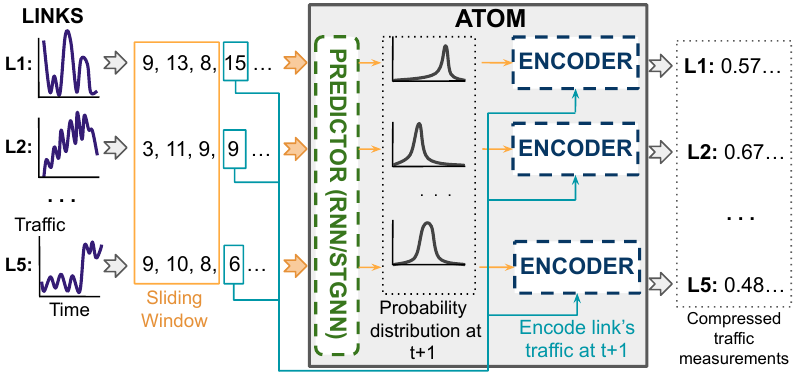}
  \caption{The predictor takes the traffic values from the SW and outputs a probability distribution per link. Then, Atom uses AC to encode the traffic value of the next time bin after the SW. 
  }
  \label{fig:archi_overview}
  \vspace{-0.3cm}
\end{figure}

\textit{Atom} takes as input the values from a single or multiple traffic measurement points, depending on the compression scenario, and outputs a floating point number. This value represents the entire sequence of compressed traffic values. In the network-wide scenario, \textit{Atom} takes as input the network topology in addition to multiple traffic values. Our compression method uses a SW to iterate over all traffic measurements and uses the traffic values within the SW to output a probability distribution. This distribution is used to actually encode the traffic measurements immediately after the SW. The process is repeated until the window has iterated over all values.

Our compression method is composed of two main blocks: predictor and encoder/decoder blocks. The predictor implements a NN-based model that predicts the probability distribution for the next time bin after the SW. The encoder/decoder takes the predicted distributions and compresses the sequences of traffic values. Figure~\ref{fig:archi_overview} shows an overview of the compressor module of \textit{Atom} with its inputs and outputs.

\subsection{Predictor}

The predictor was implemented using a Recurrent Neural Network (RNN)~\cite{10.5555/553011} when compressing the traffic values of a single measurement point. Specifically, the RNN processes the sequence of traffic values, and subsequently a Multi-Layer Perceptron (MLP) takes the resulting hidden state and outputs the mean and standard deviation of a probability distribution (e.g., Normal, Laplace). The probability distribution is then used by the AC to code the real link measurement in the next time bin. Note that the RNN architecture can only exploit temporal correlations.

In the network-wide scenario, we implemented the predictor using a custom ST-GNN. This neural architecture uses a custom message passing algorithm for each time bin to exploit both spatial and temporal correlations simultaneously. The ST-GNN uses traffic measurements and network topology as inputs. Then, the ST-GNN exchanges information between links and propagates link-level information across the topology for each time bin. Finally, the model outputs a per link probability distribution.

\begin{figure}[!t]
  \centering
  \includegraphics[width=0.99\linewidth]{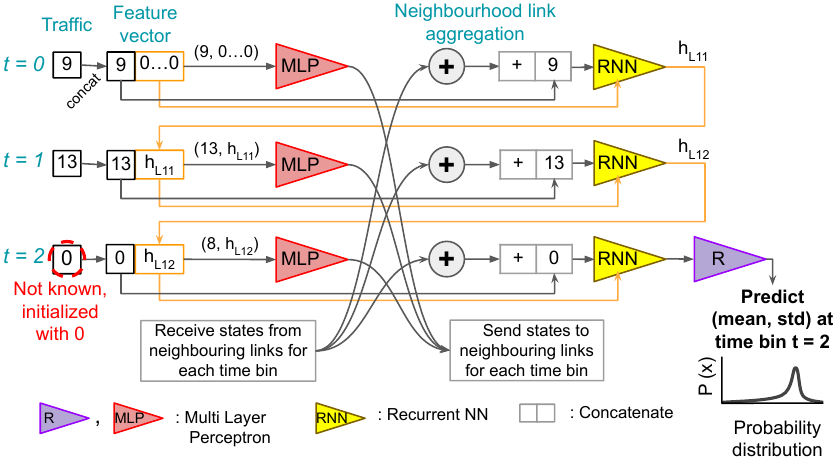}
  \caption{Overview of the ST-GNN when predicting the probability distribution for a single link at time bin \textit{t=2}. 
  }
  \label{fig:space_time_mp}
 \vspace{-0.5cm}
\end{figure}

Figure~\ref{fig:space_time_mp} shows an overview of the proposed ST-GNN architecture when predicting the probability distribution for a single link using a SW size of two. At time bin \textit{t=0}, the links' feature vector is initialized with the traffic value and padded with 0. Then, the feature vector is processed by an MLP and sent to all neighboring links, aggregating them using a sum. Simultaneously, the current link receives the hidden states from neighboring links, aggregates them using a sum, and concatenates the actual link traffic value. The resulting hidden state is then processed by an RNN, which outputs the final hidden state for the present time bin. This hidden state is used in the next time bin, \textit{t=1}, to initialize the feature vector. This process is repeated for all time bins within the SW. In the last bin, a different MLP (denoted by R in the figure) is used to process the final hidden state and outputs the parameters of the probability distribution, which is used by the AC to compress the traffic value of the current link at time bin \textit{t=2}.

\subsection{Encoder/Decoder}
\label{subsec:encoddecod}

The encoder/decoder is responsible for compressing or decompressing actual sequences of traffic values. Specifically, it uses the probability distribution from the predictor to compress/decompress the actual traffic value. We implemented the encoder/decoder using AC~\cite{10.1145/214762.214771}, compressing each traffic sequence into a single floating-point number. The compression performance of AC is heavily reliant on the predictor. A well-estimated probability distribution allows for a more efficient data representation and achieves higher compression ratios. When compressing a sequence of values, having a reliable predictor becomes crucial as it enables accurate estimation of the underlying probability distribution, capturing the temporal dependencies and patterns present in the data. 

\subsection{Mask}
\label{subsec:mask}

The ST-GNN model uses a mask to exploit the spatial correlations between links, enabling the model to learn the conditional distribution $p(x_l|\bm x_{t,<l},\bm x_{<t})$ from Equation~\ref{fig:spatial_equation}. Specifically, the mask is used to gradually incorporate the already compressed/decompressed link traffic values into the prediction. By masking the known link traffic values, our model learns to predict the conditional probability distribution to the known traffic values, thus exploiting the spatial correlations between links. 

During training, the mask of unknown links is randomly created. This means that for each SW, we associate a random mask over the links to indicate the link traffic values that are known. In the compression/decompression phase, the mask starts by marking all traffic values as unknowns. Then, the predictor compresses/decompresses the traffic values, incorporating them into the prediction by changing the mask.

\subsection{Compression/Decompression}
\label{subsec:compression}

\begin{figure}[!t]
  \centering
  \includegraphics[width=0.8\linewidth]{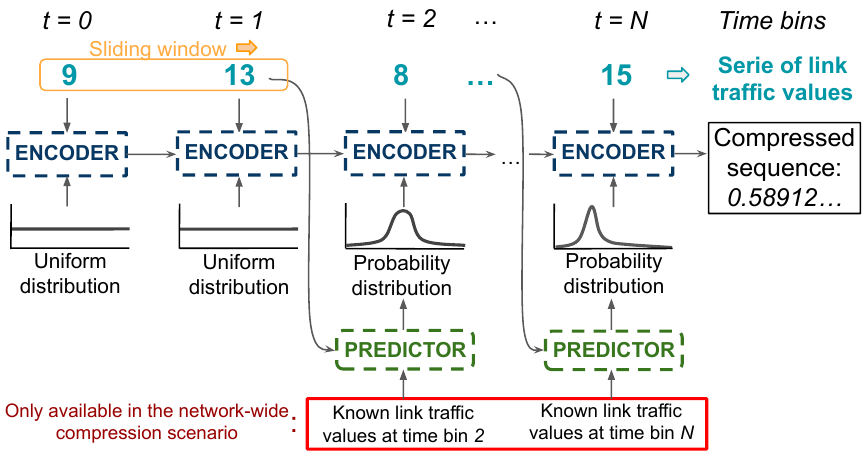}
  \caption{Overview of the compression process for a single link. In the network-wide scenario, the predictor outputs the conditional probability distribution $p(x_l|\bm x_{t+1,<l},\bm x_{<t+1})$. }
  \label{fig:encoder_framework}
   \vspace{-0.5cm}
\end{figure}

Figure~\ref{fig:encoder_framework} presents an overview of the compression process for a single link/measurement point. Recall that the predictor is implemented using an RNN for single-link compression. In the network-wide scenario, the predictor is implemented with an ST-GNN that takes the traffic values of the neighboring link as additional input features. Note that \textit{Atom} uses uniform probability distributions to encode the traffic values within the first SW.

For decompression, the model reads the compressed file and uses the same NN-based model to recover the original sequence bin by bin. Similarly to the compression phase, the first time bins are decompressed using uniform probability distributions. Then, the predictor uses the recovered values to compute the probability distributions of the links. Note that, for each time bin, the predictor receives the same input information as in the compression phase. The decoder also receives the same information; however, in this case, its operations are inverted for decoding.

\section{Experimental Results}

\begin{figure}[!t]
  \centering
  \includegraphics[width=0.99\columnwidth]{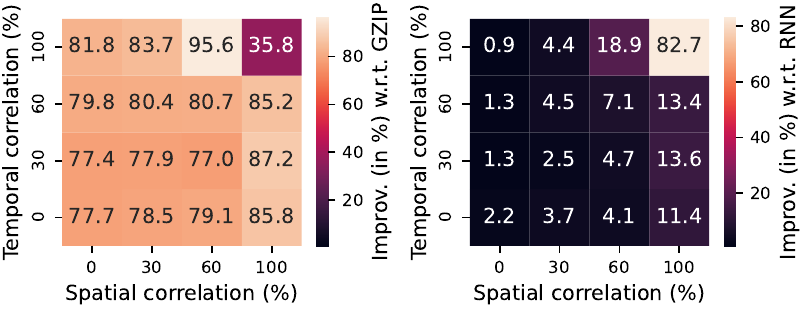}
  \vspace{-0.1cm}
  \caption{ST-GNN's compression ratio improvement (in \%) with respect to GZIP (left) and RNN (right). 
  }
  \label{fig:heatmap}
  \vspace{-0.3cm}
\end{figure}

In the first experiment, we generated synthetic datasets with different intensities of spatial and temporal correlations in the NSFNet topology \cite{Hei:04} which has 42 directional links. In addition, we evaluated the compression capabilities of \textit{Atom} on three real-world datasets. The first two datasets were the Abilene and Geant datasets with 30 and 72 directional links and with six and four months of traffic measurements respectively \cite{SNDlib10}. The third dataset was obtained from an in-house monitoring platform on a campus network. The dataset contained 12 months of traffic measurements (from December 2020 to December 2021), and the topology contained 16 directional links.

The experiments were performed on a single machine with an AMD Ryzen 9 5950X 16-Core Processor with a GeForce GTX 1080 Ti GPU to train the models. We trained all the NN-based models using 70\% of the samples for training and 30\% for evaluation. We used the negative log likelihood loss function of the Laplace distribution. For more implementation details, the source code and datasets used in this study are publicly available\footnote{\url{https://github.com/BNN-UPC/Atom_Neural_Traffic_Compression}}.

\subsection{Evaluation on synthetic data}
\label{subsec:synth_data_gen}

In the first set of experiments, we evaluated the compression performance on synthetic data originating from \textit{sine} signals. If all flows originate from the same signal, the links are highly correlated in space because their values increase or decrease proportionally for each time bin. The degree of spatial correlation in our datasets was controlled by adding random noise to the signal, shifting its phase, and changing its periodicity. In addition, to control the intensity of the temporal correlation in the experiment, we controlled the percentage of flows with added noise. We established four degrees of spatial/temporal correlations from low to high: 0\%, 30\%, 60\%, and 100\%, indicating the percentage of flows that conserved the original signal (i.e., they did not have added noise). As a result, a total of 16 experiments corresponded to all possible combinations of spatial and temporal correlation intensities, resulting in 16 ST-GNN and 16 RNN-based trained models.

Figure~\ref{fig:heatmap} (left) shows the percentage of compression ratio improvement for the ST-GNN with respect to the GZIP baseline in the network-wide scenario. The results indicate that ST-GNN outperformed GZIP by a large margin in all experiments. Note that the scenario with the maximum spatial and temporal correlations contains a small number of traffic values that are repeated frequently. GZIP uses Huffman coding~\cite{4051119} as the underlying algorithm, which can effectively exploit highly repeated traffic values, thus explaining why the compression ratio improvement is the lowest. The results also indicate the expected performance of ST-GNN when evaluated in real-world scenarios. In particular, the intensity of the temporal and spatial correlations of a real-world dataset points to the expected performance with respect to GZIP. Figure~\ref{fig:heatmap} (right) shows the performance improvement of ST-GNN models with respect to RNN-based models. The ST-GNN model exhibits remarkable performance in scenarios with high spatial correlation, indicating that it has the flexibility to exploit both spatial and temporal correlations.

\begin{figure}[!t]
\centering
    \subfigure[]
    {\includegraphics[width=0.480\columnwidth]{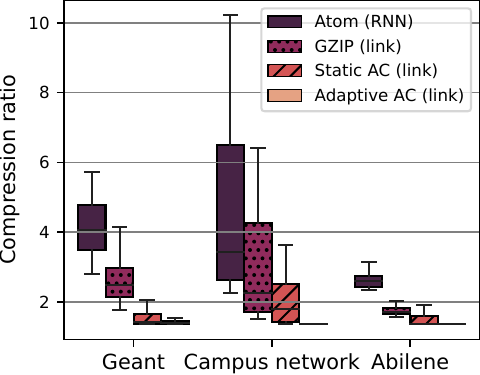}
    \label{fig:link_level_compression}}
    \subfigure[]
    {\includegraphics[width=0.480\columnwidth]{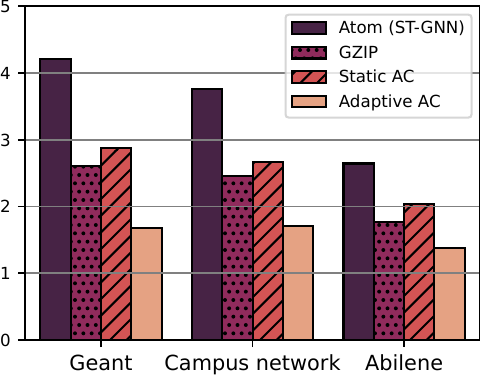}
    \label{fig:bar_plot_comp_ratios}}
    \caption{Compression ratios for the single-link (a) and for the network-wide (b) scenarios.}
    \label{fig:temp_spati_corr_evol}
\vspace{-0.5cm}
\end{figure}

\subsection{Compressing real-world data}
\label{subsec:comp_real_world}

In this set of experiments we compressed three real-world datasets and compared the compression performance against three baselines: Static AC, Adaptive AC and GZIP. Static AC iterates over the entire sequence and computes the probability distribution for all traffic values. It then compresses or decompresses the entire sequence using the same static distribution for each AC coding step. The Adaptive AC baseline computes a new distribution using the values within the SW. These baselines were intended to demonstrate the benefits of using NNs to implement the predictor module. In other words, we wanted to show that compression performance highly depends on the accuracy of the predictor. Finally, GZIP was applied to compress the entire sequence. 

Figure~\ref{fig:link_level_compression} shows the compression ratios for three real-world datasets in the first scenario. Recall that the goal is to compress the traffic values independently from multiple measurement points. Each baseline was applied to compress each sequence individually. The results indicate remarkable performance of \textit{Atom} for all datasets, outperforming GZIP by a large margin. Figure~\ref{fig:bar_plot_comp_ratios} shows the experimental results of compressing the same datasets in a network-wide scenario. In this scenario, Static AC computes the probability distribution using the entire dataset (i.e., including all links), and Adaptive AC updates the distribution by including the values from all links within the SW. The results indicate that \textit{Atom} achieved the highest compression ratios for all three datasets. In particular, it outperformed Static AC by $\approx$47\%, $\approx$41\%, and $\approx$29\% for the Geant, campus network, and Abilene datasets, respectively. In addition, the performance improvements with respect to GZIP are $\approx$62\%, $\approx$53\%, and $\approx$50\% for the same datasets. These results demonstrate the benefits of leveraging NNs to exploit spatial and temporal correlations for traffic compression.

\subsection{Cost}

\textit{Atom} compresses traffic measurements in a streaming manner as they originate from the network-monitoring platform. Conversely, GZIP must wait for the entire dataset to apply the compression algorithm. Alternatively, GZIP can compress all links simultaneously in each time bin independently. We performed this experiment, and the results indicated that GZIP achieves compression ratios of $\approx$0.94, $\approx$0.4, and $\approx$0.54 for the Geant, campus network, and Abilene datasets, respectively. In other words, the compressed data occupies more space than the original data.

We also computed the average cost of compressing a time bin using \textit{Atom} to evaluate its deployment in real-world online traffic compression. The results can be seen in Table~\ref{table:cost}, which indicates that \textit{Atom} is capable of effectively compressing 5-minute bins in the order of seconds, thereby enabling online compression. Additionally, we computed the size of the weights of the model stored in a file. This indicates that the \textit{Atom} is a lightweight method that achieves high compression ratios with an expendable model size.

\begin{table}[!t]
\centering
\begin{tblr}{
  colspec = {crrrr},
  cell{1}{2,4} = {c=2}{c}, 
}
\hline[2pt]
  Dataset &   Mean Cost (s) &      &  Model size (KB) &      \\
\hline[1pt]
       &    ST-GNN &  RNN &  ST-GNN & RNN \\
\cline{2-5}
    Geant  &  3.20 & 0.47 & 575 & 489 \\
    Campus network  & 0.22  & 0.07 & 236 & 202 \\
    Abilene  &  1.22 &  0.29 & 332 & 285 \\
\hline[2pt]
\end{tblr}
\vspace{0.1cm}
\caption{Atom's mean cost to compress a time bin (in seconds) and model size (in Kilobytes).}
\label{table:cost}
\vspace{-0.8cm}
\end{table}

\section{Related Work}

In the last years, NNs have started to be used for compressing images, videos and voice\cite{8693636}. However, in the networking field, existing compression methods follow traditional approaches, with GZIP being the most popular method for network traffic compression. The work in \cite{aiello2005sparse} proposed a lossy method to approximate real network traffic by capturing the most relevant traffic features. In \cite{beirami2015packet} the authors proposed exploiting the traffic redundancies at the packet level to reduce the transmitted traffic. In \cite{mehboob2006high} they proposed an architecture for implementing the LZ77 compression algorithm \cite{ziv1977universal} on an FPGA. The work of \cite{fusco2010net} describes a solution for on-the-fly storage, indexing, and querying of network flow data. A more recent work leverages the P4 language~\cite{bosshart2014p4} and generalized deduplication to implement a solution that operates at line-speed. To the best of our knowledge, no previous work has explored the use of NNs for the purpose of compressing network traffic data.

\section{Conclusion}

Existing methods for network traffic compression are generic, resulting in low compression ratios. This limitation becomes even more critical when traffic is compressed in an online scenario. In our work, we explored the use of NNs and arithmetic coding to compress traffic measurements. Specifically, we presented \textit{Atom}, a model that exploits the spatial and temporal correlations intrinsic to traffic measurements. In addition, we designed a custom ST-GNN to simultaneously exploit spatio-temporal correlations. The experimental results show that \textit{Atom} can effectively compress real-world traffic, with an improvement of $\geq$50\% in the compression ratio for real-world datasets with respect to GZIP. 

\section*{Acknowledgments}
This work was supported by the Spanish Ministry of Economic Affairs and Digital Transformation and the European Union – NextGeneration EU, in the framework of the Recovery Plan, Transformation and Resilience (PRTR) (Call UNICO I+D 5G 2021, refs. number TSI-063000-2021-3,  TSI-063000-2021-38, and TSI-063000-2021-52). It has also received funding from the European Union’s Horizon 2020 research and innovation program under grant agreement no. 101017109 “DAEMON”. This publication is part of the Spanish I+D+i project \mbox{TRAINER-A} (ref.~PID2020-118011GB-C21), funded by MCIN/AEI/10.13039/501100011033. It was also partially funded by the Catalan Institution for Research and Advanced Studies (ICREA) and the Secretariat for Universities and Research of the Ministry of Business and Knowledge of the Government of Catalonia and the European Social Fund.
Krzysztof Rusek was supported by the Polish Ministry of Science and Higher Education with the subvention funds of the Faculty of Computer Science, Electronics and Telecommunications of AGH University of Science and Technology.

\bibliographystyle{ACM-Reference-Format}
\bibliography{sample-base}

\end{document}